\documentclass[superscriptaddress,prd,preprintnumbers,twocolumn,nofootinbib]{revtex4}

\usepackage{amsmath,amssymb,amsfonts}
\usepackage{bm, slashed, soul}
\usepackage{color,graphicx}
\usepackage{hyperref}
\usepackage{multirow}

\usepackage{xcolor}
\definecolor{red}{rgb}{0.9, 0,0}

\newcommand{\be}{\begin{equation}}
\newcommand{\ee}{\end{equation}}
\newcommand{\bea}{\begin{eqnarray}}
\newcommand{\eea}{\end{eqnarray}}

\def\beq#1\eeq{\begin{align}#1\end{align}}
\def\beqnn#1\eeq{\begin{align*}#1\end{align*}}

\begin{document}

\title{A map of the non-thermal WIMP
}

\author{Hyungjin Kim}
\email{hjkim06@kaist.ac.kr}
\affiliation{Department of Physics, KAIST, Daejeon 34141, Korea}
\affiliation{Center for Theoretical Physics of the Universe, Institute for Basic Science (IBS), Daejeon 34051, Korea} 
\author{Jeong-Pyong Hong}
\email{hjp0731@icrr.u-tokyo.ac.jp}
\affiliation{Institute for Cosmic Ray Research, The University of Tokyo, 5-1-5 Kashiwanoha, Kashiwa, Chiba 277-8582, Japan}
\affiliation{Kavli IPMU (WPI), UTIAS, The University of Tokyo, 5-1-5 Kashiwanoha, Kashiwa, Chiba 277-8583, Japan}
\author{Chang Sub Shin}
\email{changsub.shin@apctp.org}
\affiliation{Asia Pacific Center for Theoretical Physics, Pohang 37673, Korea} 
\affiliation{Department of Physics, Postech, Pohang 37673, Korea}

\date{\today}

\begin{abstract}
We study the effect of the elastic scattering on the non-thermal WIMP, which is produced by direct decay of heavy particles at the end of reheating. 
The non-thermal WIMP becomes important when the reheating temperature is 
well below the freeze-out temperature. 
Usually, two limiting cases have been considered.  One is that the produced high energetic dark matter particles are quickly thermalized due to the elastic scattering with background radiations. 
The corresponding relic abundance is determined by the thermally averaged annihilation cross-section at the reheating temperature. The other one is that 
the initial abundance is too small for the dark matter to annihilate so that the
 final relic is determined by  the initial amount itself. We study the regions between these two limits, and show that the relic density depends not only on the annihilation rate, but also on the elastic scattering rate. 
Especially, the relic abundance of the $p$-wave annihilating dark matter crucially relies on the elastic scattering rate because the annihilation cross-section is sensitive to the dark matter velocity.
We categorize the parameter space into several regions where each region has distinctive mechanism for determining the relic abundance of the dark matter at the present Universe.
The consequence on the (in)direct detection is also studied.   
\end{abstract}

\maketitle
\preprint{APCTP Pre2016-021}
\preprint{CTPU-16-32}

\preprint{IPMU16-0164}
\section{Introduction}

The weakly interacting massive particle (WIMP) is one of the
promising dark matter (DM) candidates because 
it can be naturally incorporated in new physics beyond the Standard Model, and it gives interesting observable consequences. 

In the standard thermal history, the most important quantity to determine 
the relic density is a thermal averaged pair annihilation cross-section, $\langle \sigma_{\rm ann} v_{\rm rel}\rangle_T$. Taking ``thermal average" is justified because the elastic scattering rate between the WIMP dark matter and the background radiation is much bigger than the annihilation rate, 
so the kinetic decoupling happens well after the dark matter freeze-out \cite{Bringmann:2006mu,Visinelli:2015eka}.
Usually, the elastic scattering does not have the special role to determine the relic density, 
but it is crucial for small scale structures since the interaction suppresses the growth of the dark matter density perturbation \cite{Boehm:2000gq,Hofmann:2001bi,Loeb:2005pm,Bertschinger:2006nq}. 
When the reheating temperature is low, its effect is more interesting depending on 
whether the kinetic decoupling happens before or after the end of reheating \cite{Erickcek:2011us,Barenboim:2013gya,Fan:2014zua,Erickcek:2015jza,Choi:2015yma}. 

In this letter, we study the possibility that the relic abundance of WIMP explicitly depends on the elastic scattering rate.
In the context of self interacting dark matter, such possibility is realized 
by noting that the dominant annihilation rate from $3\to 2$ scattering and 
the elastic scattering between the DM and thermal bath can be independent so that we can take 
suitable parameters to show such behavior \cite{Kuflik:2015isi}. This assumption is not usually valid in the case of WIMP because the annihilation and elastic scattering cannot be treated independently.

However, when the reheating temperature is well below the dark matter mass, a new possibility emerges. 
At the end of reheating, the dark matter is produced by the direct decay of 
heavy particles. Such non-thermally produced dark matter particles have very high energies. 
Evolution of the dark matter momentum gives a strong effect on the annihilation cross-section, and such evolution is determined by the elastic scattering rate. 
Consequently, the relative size of the annihilation rate, 
the elastic scattering rate, and the Hubble rate at the end of reheating can 
give various mechanisms to determine the final relic density of the dark matter.  
We classify the parameter space into the regions where each region has distinctive mechanism to determine the relic density of the dark matter. We also provide analytic expressions and numerical results for each of those mechanisms.
Especially, we find that the $p$-wave annihilating dark matter has more interesting property because the cross-section highly depends 
on the expectation value of the dark matter momentum.

In section \ref{sec:TH}, we present our basic set-up. 
In section \ref{sec:Eofmo}, we compute the momentum evolution of the dark matter after its production at the end of reheating.
The effect of the momentum evolution on the annihilation cross-section and the corresponding final abundance of the dark matter are discussed in section \ref{sec:Eofann}. We discuss the constraints from (in)direct detection experiments in section \ref{sec:Cons}, and conclude in section \ref{sec:Conc}.

\section{Thermal History of the Non-thermal WIMP}\label{sec:TH}

In our set-up, there is the early stage of the matter dominated Universe maintained by a long lived heavy particle, $\phi$. 
After most of  $\phi$ decay, the Universe is ``reheated" and radiation ($\gamma$) starts to dominate the energy density of the Universe with a reheating temperature, $T_{\rm reh} \sim \sqrt{\Gamma_\phi M_{Pl}}$.
On one hand, the dark matter ($\chi$) can be produced either from the scattering of the radiation background, or from the direct decay of $\phi$ with a branching fraction ${\rm Br}_\chi$.

Ignoring the sub-leading contributions, the corresponding Boltzmann equations of each components are given as 
\beq
\dot\rho_\phi =&-3 H \rho_\phi -\Gamma_\phi \rho_\phi, \nonumber\\
\dot\rho_\gamma  = & - 4 H \rho_\gamma+ {\rm Br}_\gamma \Gamma_\phi \rho_\phi,\nonumber\\
\dot n_\chi = &-3 H n_\chi  +{\rm Br}_\chi \Gamma_\phi \frac{\rho_\phi}{m_\phi} -\langle \sigma_{\rm ann} v_{\rm rel}\rangle_\chi  n_\chi^2,\nonumber\\
&  + \langle\sigma_{\rm ann} v_{\rm rel}\rangle_T (n_\chi^{\rm eq})^2,\nonumber\\
H=& \sqrt{\frac{\rho_\phi + \rho_\gamma + \rho_\chi}{3M_{Pl}^2}}, 
\eeq  where $M_{Pl}$ is the reduced Planck mass.
Here we consider a situation that 
 the reheating temperature is lower than the thermal freeze-out temperature of the dark matter ($T_{\rm fr}$). Before the end of the reheating, $\Gamma_\phi \leq H$,
there are several sources for the dark matter density. 
First of all,  a usual  freeze-out mechanism can work with 
$n_\chi |_{T_{\rm fr}} \sim H(T_{\rm fr})/\langle\sigma_{\rm ann} v_{\rm rel}\rangle_{T_{\rm fr}}$, while the resulting abundance is subsequently diluted by continuous entropy injection. 
If ${\rm Br}_\chi$ is big,  quasi-static equilibrium state can persist until the end of reheating ($\langle\sigma_{\rm ann} v_{\rm rel}\rangle_T n_\chi^2 \sim {\rm Br}_\chi \Gamma_\phi \rho_\phi/m_\phi$) \cite{Cheung:2010gj}. Also  if $m_\phi$ is large enough to satisfy $m_\phi \gtrsim m_\chi^2/T$, production from 
an inelastic scattering between  thermal bath and 
a boosted radiation produced by $\phi$ decays becomes important \cite{Allahverdi:2002pu,Harigaya:2014waa}.
Here we take a rather moderate hierarchy between the mass of $\phi$ and $\chi$ as $m_\phi/m_\chi={\cal O}(10-100)$, and a sub GeV reheating temperature so that the inelastic scattering is subdominant. 
For $T_{\rm reh}\ll m_\chi$ and a sizable ${\rm Br}_\chi$, the most important source of the late time dark matter abundance is the direct decay of $\phi$ at the end of reheating. 
It is known that such  non-thermal production of the DM can be simplified by assuming that the dark matter is instantaneously produced from the heavy particle decay at $T=T_{\rm reh}$ with an initial amount of the DM given as $n_\chi^{\rm reh}= {\rm Br}_\chi \rho_\phi/m_\phi$ \cite{Choi:2008zq,Acharya:2009zt,Baer:2014eja,Kane:2015qea}.
In summary, we are interested in the following range of parameters:
\beq
T_{\rm reh}\leq T_{\rm fr} \ll m_\chi \ll m_\phi \sim {\cal O}(10-100)m_\chi.
\eeq
Because of the hierarchy between masses, the initial energy of the DM is much greater than $m_\chi$. So we first consider the evolution of the dark matter momentum and then consider its effect on the annihilation rate.

\section{Evolution of the DM momentum}\label{sec:Eofmo}

After the dark matter is produced, it experiences two types of interactions. One is the elastic scattering by the background radiations ($\chi\gamma\to \chi\gamma$). The other one is pair annihilation of the dark matter into the radiations ($\chi\chi\to \gamma\gamma$). The effect of pair production from thermal bath $(\gamma\gamma\to \chi\chi)$ is negligible if the DM abundance at $T_{\rm reh}$ is much larger than the equilibrium value. 
 In principle, both of the elastic scattering and annihilation are relevant for the dark matter momentum evolution. However, as we discuss in appendix~\ref{apdix}, the contribution from annihilation can be safely ignored when the momentum distribution of $\chi$ has a narrow width compared to its mean value. 
Initially, dark matters are produced from the decays of $\phi$, so the width $\sigma_{sd,\chi} \sim \Gamma_\phi$ is naturally smaller than the mean value of momentum, $\langle p\rangle_\chi\sim m_\phi$. Until the relaxation time when the momentum distribution arrives at its equilibrium one up to the normalization factor, the width of the distribution is still small and our simplification is justified.
Then the momentum evolution is governed by the following equation~\cite{Hisano:2000dz},
\beq\label{eq:momentum_evol}
 \frac{d  p_\chi }{dt } +  H   p_\chi   &=
-n_\gamma
\bigg\langle \int d(\sigma_{\rm el}v_{\rm rel}) \Delta p
\bigg\rangle_{\chi,T}  \nonumber\\
&\equiv
 - n_\gamma\langle \sigma_{\rm el} v_{\rm rel}  
\Delta p_\chi\rangle_{\chi,T}\, ,
\eeq 
where  $p_\chi \equiv \langle p \rangle_\chi$,    
$\langle\cdots \rangle_{\chi, T}$ 
is the average over the distribution of $\chi$ and $\gamma$ 
in the rest frame of the thermal plasma,  $\sigma_{\rm el}$ is the elastic scattering cross-section,
$\Delta p $ is the change of the dark matter momentum from single event of the elastic scattering, and $n_\gamma$ is the number density of the background radiation, $n_\gamma\sim g_* T^3$.

 The right hand side of \eqref{eq:momentum_evol} can be simplified as $p_\chi n_\gamma$ times
  \beq
  \frac{ \left\langle \sigma_{\rm el} v_{\rm rel}  \Delta p_\chi \right\rangle_{\chi, T}  }{p_\chi} 
    \simeq\left\{\begin{array}{ll} \langle \sigma_{\rm el} v_{\rm rel}\rangle_{\chi, T}   & 
 (I) , \\ 
  \langle \sigma_{\rm el} v_{\rm rel}\rangle_{\chi, T}\frac{p_\chi T}{m_\chi^2}   & 
   (II) , \\
  \langle\sigma_{\rm el}v_{\rm rel}\rangle_{\chi, T} \frac{T}{m_\chi} 
  \left(1 - \frac{3 m_\chi T}{p_\chi^2}\right)  & 
  (III)  ,\end{array}\right.
 \eeq 
 for different ranges of the dark matter momentum,
 \beq
 (I)&\  m_\chi^2 \ll p_\chi T, \nonumber\\
(II)&\  p_\chi T \ll m_\chi^2 \ll p_\chi^2, \nonumber\\
(III)&\  p_\chi T \ll p_\chi^2 \ll m_\chi^2.
 \eeq
In ($I$), the DM is relativistic in the plasma rest frame, and in the center of mass (cm) frames. 
In ($II$), the DM is non-relativistic in the cm frame, whereas still relativistic in the plasma rest frame. 
In ($III$), the DM is non-relativistic in both frames. 
In the last case, the additional factor in the elastic scattering rate drives $p_\chi$ to the equilibrium value, $p_\chi^{\rm eq}=\sqrt{3m_\chi T}$. 
 For the order of magnitude estimation, we obtain
$\langle \sigma_{\rm el} v_{\rm rel} \Delta p_\chi\rangle/p_\chi
\sim \langle\sigma_{\rm el} v_{\rm rel}\rangle \langle\Delta p\rangle_\chi/p_\chi $. 
Thus the additional factors of the scattering rate,
$1\, (I)$, $p_\chi T/m_\chi^2\,(II)$, $T/m_\chi \,(III)$  are easily understood from the fact that the allowed phase space, $ \langle\Delta p\rangle_\chi/p_\chi$,  becomes wider as the collision energies become higher. 
Since the common factor $n_\gamma$ depends on the temperature, the DM can quickly arrive at kinetic equilibrium or it can just decouple relativistically depending on $T_{\rm reh}$.  

\begin{figure}[t]
\begin{center}
\includegraphics[width=8.5cm]{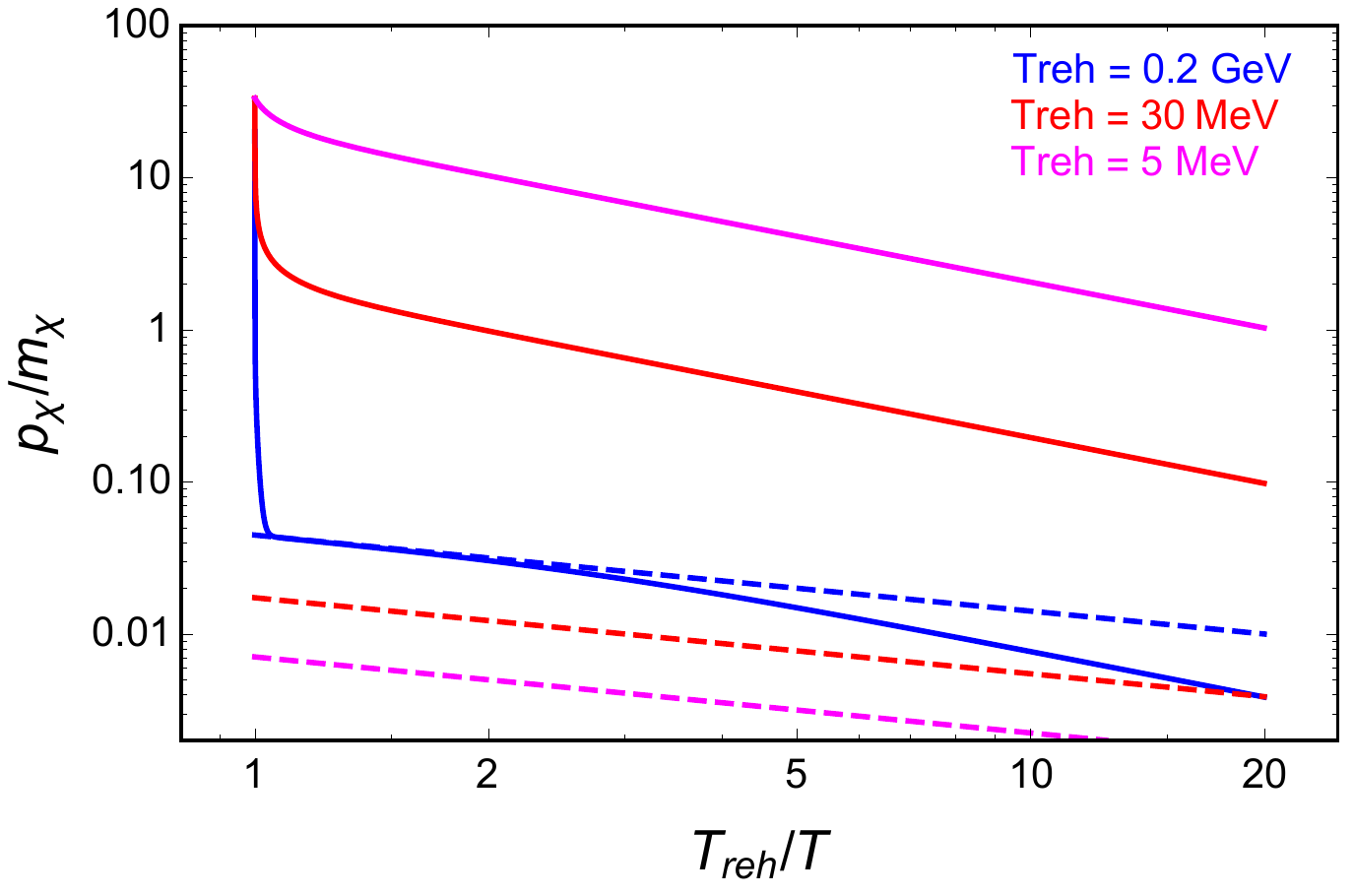}
\caption{Momentum evolution of  the non-thermally produced dark matter for different reheating temperatures, $T_{\rm reh} = 0.2$\,GeV\,(Blue), 30\,MeV\,(Red), 5\,MeV\,(Magenta).  In this plot, we set $m_\chi = 300\,\text{GeV}$. The initial momentum of the DM is given as $p_{\chi}^{\rm reh} = 20\, m_\chi$. The dashed lines are the kinetic equilibrium values, $p_\chi^{\rm eq} = \sqrt{3 m_\chi T}$.  }     
\label{fig:momentum_evol}    
\end{center}
\end{figure}
When the initial momentum of the dark matter is much greater than $m_\chi$,
 Fig.~\ref{fig:momentum_evol} shows possible evolution of the momentum for different reheating temperatures. 
 At a relatively high reheating temperature,  the elastic scattering rate is large enough to make 
 the dark matter in kinetic equilibrium instantaneously after its production. 
 As the temperature goes down, the momentum follows the equilibrium value ($
 p_\chi^{\rm eq}  \propto 1/ \sqrt{a}$) 
 until the kinetic decoupling. 
If the reheating temperature is relatively low,  
after the dark matter momentum experiences a small sharp suppression around $T_{\rm reh}$, 
it slowly decreases as $p_\chi\propto 1/a$, 
and  it could become non-relativistic well after reheating (magenta line).

There is a natural connection between the momentum evolution and the dark matter pair annihilation rate.  
For example, if they are highly relativistic, the cross-section becomes $\langle \sigma_{\rm ann} v_{\rm rel}\rangle_\chi \propto 1/p_\chi^2$, and it will increase as the energy of the particle decreases. 
Therefore, if the dark matter is not instantaneously thermalized, the annihilation of the dark matter could happen later when the annihilation cross-section becomes large enough to start the annihilation. 

Since the corresponding dark matter abundance is affected by the evolution of the annihilation cross-section, we can find the connection between the final yield of the dark matter, and the elastic scattering rate.

\section{Evolution of the pair annihilation cross-section}\label{sec:Eofann}

When the initial abundance ($n_{\chi}^{\rm reh}$) is much greater than  $n_{\chi}^{\rm eq}$, 
the production of $\chi$ from thermal bath can be ignored, and
the corresponding Boltzmann equation for the 
dark matter number density is simplified as
\beq
\dot n_\chi + 3 H n_\chi =  -\langle\sigma_{\rm ann} v_{\rm rel}\rangle_{\chi}  n_\chi^2. 
\eeq
Solving the above equation, we find the yield of the dark matter at the present time, $t_0 \gg t_{\rm reh}$, as
\beq\label{eq:yield}
Y_\chi(t_0) =& Y_\chi(t_{\rm reh})
\left(1+  \frac{n^{\rm reh}_\chi}{H_{\rm reh}}\int_0^1  d u
 \langle \sigma_{\rm ann} v_{\rm rel}\rangle_{\chi}  \right)^{-1},\eeq
where  $u\equiv  \sqrt{t_{\rm reh}/t}$,
 $H_{\rm reh}$ is the Hubble rate, and $s_{\rm reh}$ is the entropy of the Universe at $T=T_{\rm reh}$.
 The yields are denoted by $Y_\chi(t_0) = (n_\chi/s)_{t_0}$, $Y_\chi(t_{\rm reh}) = n_\chi^{\rm reh}/s_{\rm reh}$.
The time dependence of $\langle \sigma_{\rm ann} v_{\rm rel}\rangle_\chi$ 
is determined by that of $p_\chi(u)$ governed by Eq.~(\ref{eq:momentum_evol}). 
More precisely, we have to evaluate the annihilation cross-section that is averaged over the full time dependent momentum distribution function of dark matter. However, when dark matters are non-thermally produced by two-body decays, the width of the distribution would be small, and for $\Delta t < 1/\Gamma_{\rm el}$,  $
\langle p^n\rangle_{\chi} = \langle p\rangle^n_\chi(1+ {\cal O}(\sigma_{sd,\chi}^2/p_\chi^2))\simeq p_\chi^n $. Thus it is a good approximation to take
$\langle \sigma_{\rm ann} v_{\rm rel}\rangle_\chi$ as the function of $p_\chi(u)$ until the relaxation time, $\Delta t \sim 1/\Gamma_{\rm el}$.
After relaxation, 
the momentum distribution will be proportional to the equilibrium value, in which 
the standard deviation and mean value are in the same order. This leads to ${\cal O}(1)$ difference between $\langle p^n\rangle_\chi$ and $\langle p\rangle_\chi^n$, but this does not change our result qualitatively. 
Solving the full Boltzmann equations will be discussed in future work.

Two limiting cases are familiar. One is 
that $p_\chi(u)$ quickly arrives at its equilibrium value within the period much
shorter than the Hubble time as given in Fig.~\ref{fig:momentum_evol} with blue color.
The dark matter annihilation happens after its thermalization but still much faster than 
the Hubble expansion rate. Therefore, 
 $Y_\chi(t_0) = H_{\rm reh}/(\langle \sigma_{\rm ann} v_{\rm rel}\rangle_{T_{\rm reh}}  s_{\rm reh})$.
The other limit is that the initial abundance is too small so that $\langle\sigma_{\rm ann} v_{\rm rel}\rangle n_\chi^{\rm reh} \ll H_{\rm reh}$. Annihilation barely happens,  and the yield is preserved;
$ Y_\chi(t_0) = Y_\chi(t_{\rm reh})$. 
In both cases, the final yields do not explicitly depend on the elastic scattering cross-section. 

There is an intermediate domain between these two limiting cases. Including the above examples, we identify three mechanisms for the relic density of the DM. 
After the production of the DM from the direct decay of the heavy particles, the relic abundance is determined by one of the following mechanisms:
\begin{itemize} 
\item (N.A.) No Annihilation: the annihilation rate, $\Gamma_{\rm ann}(T, E_\chi)$, is always smaller than $H(T)$ for $T\leq T_{\rm reh}$, regardless of the dark matter momentum. Therefore the dark matter does not annihilate after the reheating, and the yield is preserved.

\item (I.A.) Instantaneous Annihilation: the elastic scattering rate, $\Gamma_{\rm el}(T_{\rm reh}, E_\chi)$, is always greater than $H_{\rm reh}$, so that 
the momentum of the dark matter quickly approaches to the equilibrium value, and 
most of the DM pair annihilation also happens at $T\simeq T_{\rm reh}$. 
Especially for the $p$-wave annihilating dark matter, the final abundance depends on the relative size of $\Gamma_{\rm ann}$ and $\Gamma_{\rm el}$.

\item (C.A.) Continuous Annihilation: 
the elastic scattering rate  becomes smaller than $H_{\rm reh}$ at $E_\chi \gg m_\chi$, so the  dark matter decouples with a relativistic energy, and travels freely after its production. 
 In this case, $\langle \sigma_{\rm ann} v_{\rm rel} \rangle \propto 1/p_\chi^2\propto a^2$, 
while $H\propto T^2\propto1/a^2$. Therefore
the annihilation could happen continuously until the DM becomes non-relativistic. 
\end{itemize}
\begin{figure}[t]
\begin{center}
\includegraphics[width=8.5cm]{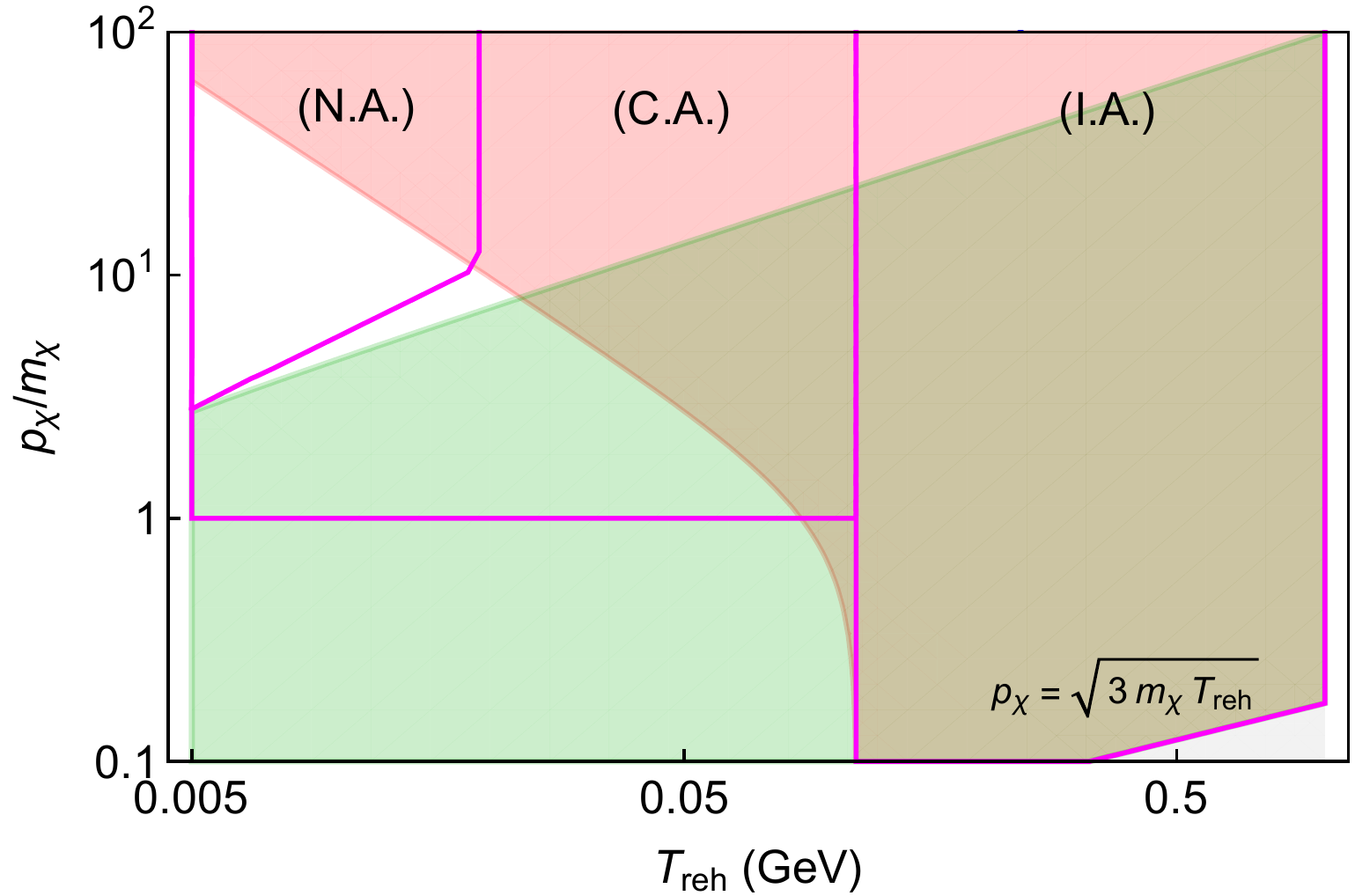}   
\caption{Illustrative plot showing the domains of different mechanism to determine the relic density of the WIMP dark matter. 
The elastic scattering is active in the red colored area: $\Gamma_{\rm el}(T_{\rm reh}, E_\chi) > H_{\rm reh}$. 
The dark matter pair annihilation is active in the green colored  area: $\Gamma_{\rm ann}(T_{\rm reh}, E_\chi) > H_{\rm reh}$.  The magenta lines are the boundary of the following three regions.
Region (N.A.): no annihilation after direct production at $T=T_{\rm reh}$. 
Region (I.A.): instantaneous thermalization and annihilation. 
Region (C.A.): the elastic scattering becomes inactive when the dark matter is relativistic, but still the pair annihilation happens after reheating.}
\label{fig:annihilation_domain} 
\end{center}
\end{figure}
The rates are given by
\beq
\Gamma_{\rm ann}(T, E_\chi ) &= n_\chi \langle\sigma_{\rm ann} v_{\rm rel}\rangle_\chi , \nonumber\\
\Gamma_{\rm el}(T, E_\chi)&= n_\gamma \frac{\left\langle \sigma_{\rm el} v_{\rm rel} \Delta p_\chi\right\rangle_{\chi,T}}{p_\chi} .
\eeq
 where $E_\chi = (m_\chi^2 + p_\chi^2)^{1/2}$.

Fig.~\ref{fig:annihilation_domain} shows the corresponding domains, heuristically.  
The red color denotes the region where the elastic scattering rate is greater than the Hubble parameter for given $p_\chi^{\rm reh}$ and $T_{\rm reh}$.
In this region, the dark matter momentum evolves nearly along the vertical direction. It quickly approaches to $p_\chi^{\rm dec}$,  which is defined as $\Gamma_{\rm el}(T_{\rm reh}, E_\chi^{\rm dec} )= H_{\text{reh}}$. If $p_\chi^{\rm dec} > p_\chi^{\rm eq}$, then the dark matter momentum redshifts as $p_\chi \propto T$.
The region (C.A.) is bounded from below by the condition that the dark matter is decoupled with a relativistic energy. 

For each regions, we can obtain the approximate formula for the final yield value by solving the Eq.~(\ref{eq:yield}). 
Since we are interested in the case where $p_\chi T_{\rm reh} < m_\chi^2$, we parameterize the annihilation and elastic scattering cross-sections in the following way:
\beq\label{eq:scat_op}
\langle\sigma_{\rm ann} v_{\rm rel}\rangle_\chi& = \frac{\alpha_{\rm ann}^2}{E_\chi^2} \left(\frac{2p_\chi^2}{E_\chi^2}\right)^{k_{\rm ann}},
\nonumber\\
 \langle \sigma_{\rm el} v_{\rm rel}\rangle_{\chi, T} &=  \frac{\alpha_{\rm el}^2}{m_\chi^2} 
\left(\frac{ E_\chi^2 T^2}{m_\chi^4}\right)^{k_{\rm el}}.
\eeq
 $k_{\rm el}$ and $k_{\rm ann}$ are the integers determined by the nature of the interactions, such as spin of the initial and final particles, and CP violating effects, etc.
 When the dark matter is non-relativistic, 
for $k_{\rm ann}=0$, the $s$-wave annihilation dominates. For $k_{\rm ann}=1$, the $p$-wave annihilation dominates. 
It is common that  $k_{\rm el}=1$ for  the elastic scattering. If the
elastic scattering is mediated by a vector boson, $k_{\rm el}= 0$ is also possible. 
In this paper, we focus on the cases with $k_{\rm el}=1$ and $k_{\rm ann}=0,1$.

\begin{figure}[t]
  \begin{center}
  \includegraphics[width=8.4cm]{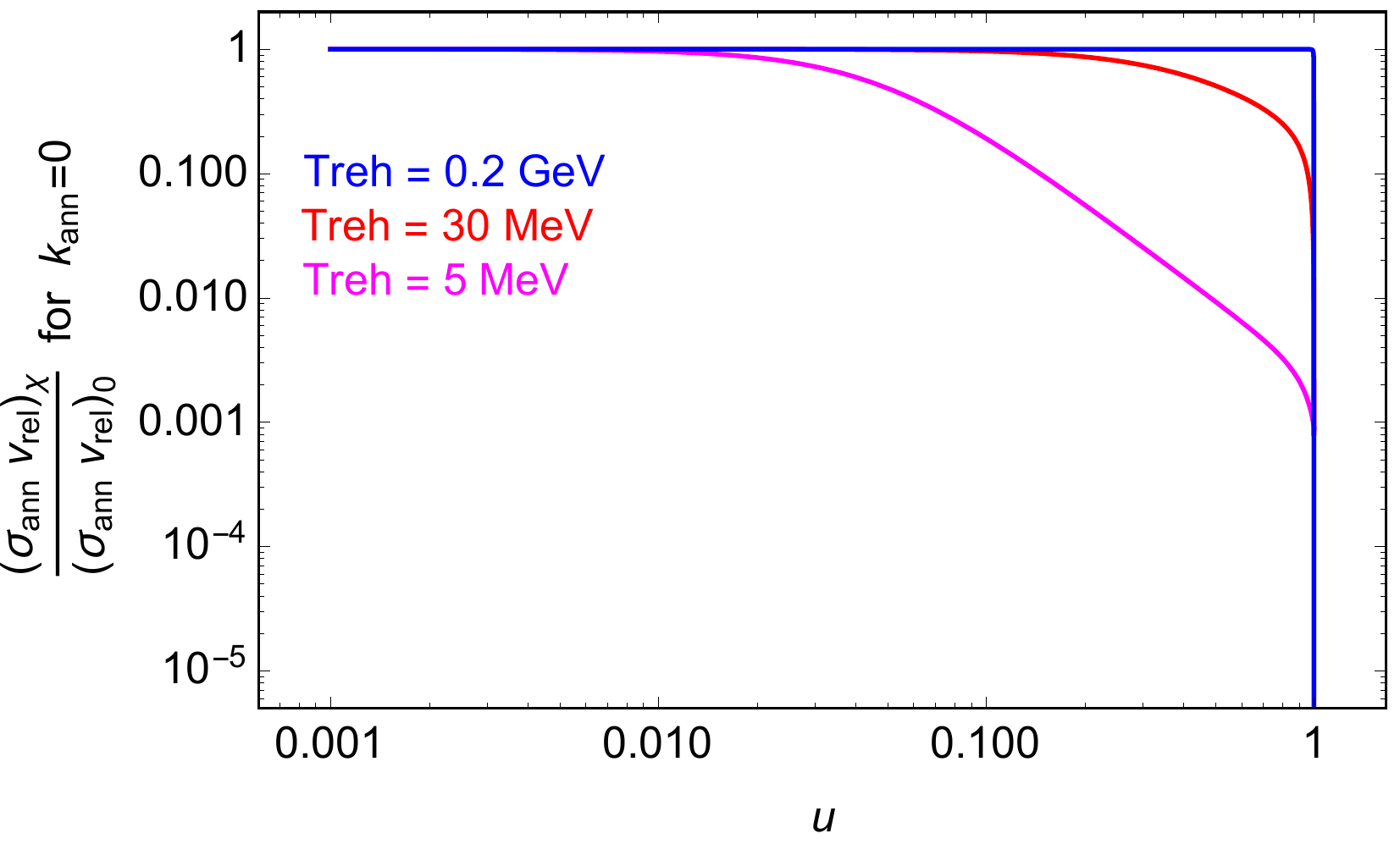} ~
 \includegraphics[width=8.5cm]{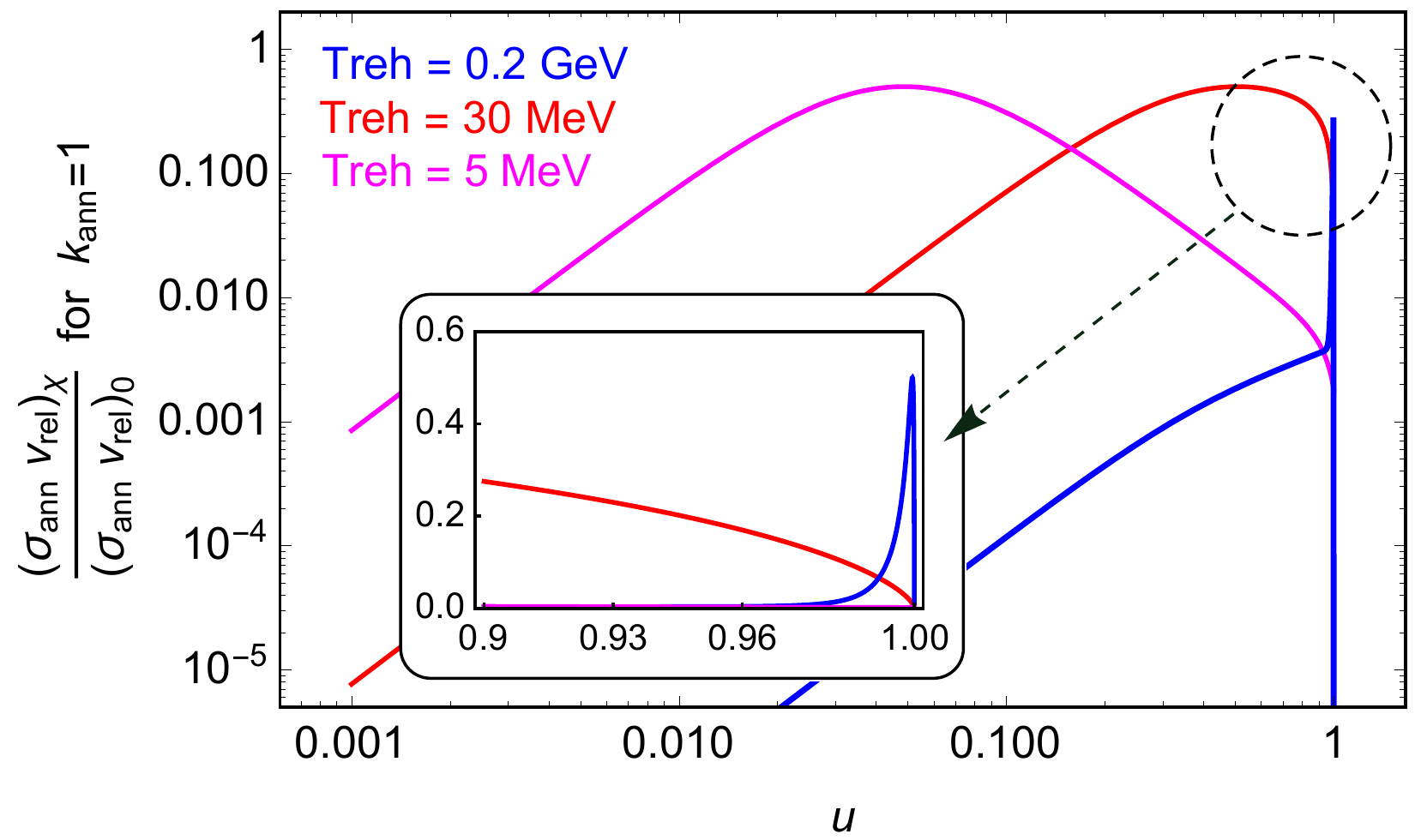}         \end{center}
    \caption{Time dependence of $\langle\sigma_{\rm ann} v_{\rm rel}\rangle_\chi/\langle\sigma_{\rm ann} v_{\rm rel}\rangle_0$ ($k_{\rm ann}=0,\,1$) 
    for the momentum evolution described in Fig.~\ref{fig:momentum_evol}. 
    For a slow varying $g_*$, $u\simeq T/T_{\rm reh}$. For $T_{\rm reh} = 0.2$\,GeV, the kinetic decoupling temperature is 
    about $T_{\rm kd}\simeq T_{\rm reh}/3$.   }
  \label{fig:annihilation_evol}
\end{figure}

Before moving forward, let us define useful quantities that are 
independent of the dark matter momentum;
\beq\label{eq:building_block}
\langle\sigma_{\rm ann} v_{\rm rel}\rangle_0 &\equiv  \frac{\alpha_{\rm ann}^2}{m_\chi^2},\ \ 
\langle\Gamma_{\rm ann}\rangle_0 \equiv  \langle\sigma_{\rm ann} v_{\rm rel}\rangle_0 n_\chi^{\rm reh},\nonumber\\
\langle\sigma_{\rm el} v_{\rm rel}\rangle_0 &\equiv  \frac{\alpha_{\rm el}^2 T_{\rm reh}^2}{m_\chi^4},\ 
\langle\Gamma_{\rm el}\rangle_0 \equiv  \langle\sigma_{\rm el} v_{\rm rel}\rangle_0 \frac{T_{\rm reh} n_\gamma^{\rm reh}}{m_\chi}.
\eeq 
In order to obtain the final yield value, the Eq.~(\ref{eq:yield}) should be evaluated. 
The integral part of Eq.~(\ref{eq:yield}) can be written as
\beq\label{eq:integral_yield}
\frac{\langle\Gamma_{\rm ann}\rangle_0}{H_{\rm reh}}
\int_0^1 du \frac{\langle\sigma_{\rm ann} v_{\rm rel}\rangle_\chi}{\langle\sigma_{\rm ann} v_{\rm rel}\rangle_0}.
\eeq
In a naive estimation, comparing $\langle\Gamma_{\rm ann}\rangle_0$ with $H_{\rm reh}$ is the only important criterion. 
Fig.~\ref{fig:annihilation_evol} shows the time dependence of the integrand, 
$\langle\sigma_{\rm ann} v_{\rm rel}\rangle_\chi/\langle\sigma_{\rm ann} v_{\rm rel}\rangle_0$.
For $k_{\rm ann}=0$, 
the annihilation cross-section approaches to $\langle \sigma_{\rm ann} v_{\rm rel}\rangle_0$ as the momentum of the dark matter decreases.
However,  for $k_{\rm ann} = 1$, there is a sharp peak around $T\simeq T_{\rm reh}$ in the region (I.A.), whose the height is $1/2$ and the width is $\Delta u \simeq H_{\rm reh}/\langle\Gamma_{\rm el}\rangle_0\ll 1$.  
Therefore, its contribution to the Eq.~(\ref{eq:integral_yield}) is of ${\cal O}(\Gamma_{\rm ann}/\Gamma_{\rm el} )$.
A simple interpretation is as follows. 
If the elastic scattering rate is large enough, the dark matter is quickly thermalized before the dark matter starts to annihilate, 
so that the peak contribution is small, and most of annihilation happens with a thermal averaged annihilation cross-section as
\beq
Y_\chi(t_0) \sim &
\frac{H_{\rm reh}}{\langle \sigma_{\rm ann} v_{\rm rel} \rangle_T s_{\rm reh}} \nonumber\\
=& \frac{H_{\rm reh}}{\langle \sigma_{\rm ann} v_{\rm rel} \rangle_0 s_{\rm reh}} \frac{m_\chi}{6T_{\rm reh}}.
\eeq
In the opposite limit, large pair annihilation can happen before the dark matter is completely thermalized. 
The corresponding yield is dominantly determined by the peak contribution as 
\beq
Y_\chi(t_0) \sim&
\frac{\langle\Gamma_{\rm el}\rangle_0}{\langle\Gamma_{\rm ann}\rangle_0} Y_\chi(t_{\rm reh})\nonumber\\
=&\frac{\langle \Gamma_{\rm el} \rangle_0}{\langle \sigma_{\rm ann} v_{\rm rel} \rangle_0 s_{\rm reh}}.
\eeq

If $\ H_{\rm reh}\gg \langle\Gamma_{\rm el}\rangle_0$, 
the production mechanism  is lying in either the domain (N.A.) or (C.A.) with the yield value,
\beq\label{eq:yield_CA}
Y_\chi(t_0) \simeq {\rm min}\left[ 
  Y_\chi(t_{\rm reh}) ,\  
    \frac{H_{\rm reh}}{\langle\sigma_{\rm ann} v_{\rm rel}\rangle_0 s_{\rm reh}} \left(\frac{c_0 H_{\rm reh}}{\langle\Gamma_{\rm el}\rangle_0}\right)^{1/3} \right],
\eeq 
where $c_0$ is an ${\cal O}(1)$ numerical constant.
The enhancement factor $(H_{\rm reh}/\langle\Gamma_{\rm el}\rangle_0)^{1/3}$ is interpreted as 
$E_\chi^{\rm dec}/m_\chi$, where $E_\chi^{\rm dec}$ is the decoupling energy at $T_{\rm reh}$. 
The reason of this factor is that the number density is mostly determined 
by $\langle\sigma_{\rm ann} v_{\rm rel}\rangle_0 n_\chi = H(T_*)$, where $T_*$ is the temperature at which 
the dark matter becomes non-relativistic. 

If  $H_{\rm reh}\ll \langle\Gamma_{\rm el}\rangle_0$, the dark matter is completely thermalized at $T_{\rm reh}$, 
and the yield is specified by either (N.A.) or (I.A.).  
For $k_{\rm ann}=0$, the yield is simply 
\beq
Y_\chi(t_0) ={\rm min}\left[ Y_\chi(t_{\rm reh}),\ 
  \frac{H_{\rm reh}}{\langle\sigma_{\rm ann} v_{\rm rel}\rangle_0 s_{\rm reh}} \right] .
\eeq
However, for $k_{\rm ann}=1$, the formula is rather complicated because the annihilation rate is highly sensitive to the momentum evolution even for the non-relativistic dark matter. The yield value is
\beq \label{eq:yield_IA}
&Y_\chi(t_0) \simeq 
{\rm min}\left[ Y_\chi(t_{\rm reh}),\right. \nonumber\\
&\left.
\frac{H_{\rm reh}}{\langle\sigma_{\rm ann} v_{\rm rel}\rangle_0 s_{\rm reh}}
\left[\frac{c_0 H_{\rm reh}}{ \langle\Gamma_{\rm el}\rangle_0} + \left(3 -\frac{T_{\rm kd}^2}{T_{\rm reh}^2}\right) \frac{T_{\rm reh}}{m_\chi}  \right]^{-1}\right].
\eeq
In the expression, the contribution of ${\cal O}(H_{\rm reh}/\langle\Gamma_{\rm el}\rangle_0)$ is coming from the peak around $u\simeq 1$. 
This also can be rephrased in terms of the kinetic decoupling temperature. After $\Delta t =1/\langle\Gamma_{\rm el}\rangle_0$, 
the elastic scattering rate scales as $\Gamma_{\rm el} \propto T^6$, while
the Hubble rate scales as $H\propto T^2$. Therefore, from $H(T_{\rm kd}) = \Gamma_{\rm el}(T_{\rm kd})$, we find
\beq
\frac{H_{\rm reh}}{\langle\Gamma_{\rm el}\rangle_0}\simeq \left(\frac{T_{\rm kd}}{T_{\rm reh}}\right)^4.
\eeq
As the reheating temperature is lower, the peak contribution becomes more important because $T_{\rm kd}$ is nearly independent of $T_{\rm reh}$.
The remaining contribution of ${\cal O}(T_{\rm reh}/m_\chi)$ is for  $u<1-\Delta u$, which gives  $\langle \sigma_{\rm ann} v_{\rm rel}\rangle_{T_{\rm reh}} n_\chi \sim  H_{\rm reh}$ if the peak contribution is neglected.

The analytic formulae are matched with each other at a naive boundary between (C.A.) and (I.A.), $c_0 H_{\rm reh} =  \langle\Gamma_{\rm el}\rangle_0$. The value $c_0$ is numerically determined to be $c_0\simeq 0.4$, as it is shown in Fig.~\ref{fig:DM_density_p}.

\section{Dark matter Constraints}\label{sec:Cons}

\subsection{Relic density}

Now we try to fit the above results to the present dark matter relic abundance \cite{Ade:2015xua}, 
\beq
\Omega_{\chi} h^2=&0.11\left(\frac{m_\chi}{100\,\rm GeV}\right) \left(\frac{Y_\chi (t_0)}{4 \times 10^{-12}}\right),
\eeq
for $\alpha=\alpha_{\rm ann}=\alpha_{\rm el}$, and for different choices of $m_\chi$ and $T_{\rm reh}$.
For the WIMP dark matter, taking $\alpha_{\rm ann}=\alpha_{\rm el}$ is a reasonable assumption. 
$\alpha$ has an upper bound from unitarity and perturbativity condition. 
Here we  take $\alpha < 1$ as the criterion for  both conditions. 
The initial yield $Y_\chi(t_{\rm reh})$ also has an upper bound.  
The direct production from heavy particle decays gives $E_\chi^{\rm reh} n_\chi^{\rm reh}= {\rm Br}_\chi \rho_\phi^{\rm reh}$, so that
\beq
Y_\chi(t_{\rm reh}) = \frac{3g_*(T_{\rm reh})}{4g_{*S}(T_{\rm reh})} \frac{{\rm Br}_\chi}{{\rm Br}_\gamma}\frac{ T_{\rm reh}}{E_\chi^{\rm reh}}.\eeq
For ${\rm Br}_\chi \lesssim {\rm Br}_\gamma$,  $Y_\chi(t_{\rm reh})$ is bounded by $T_{\rm reh}/E_\chi^{\rm reh}$.

\begin{figure}[t]
  \begin{center}
   \includegraphics[width=8.2cm]{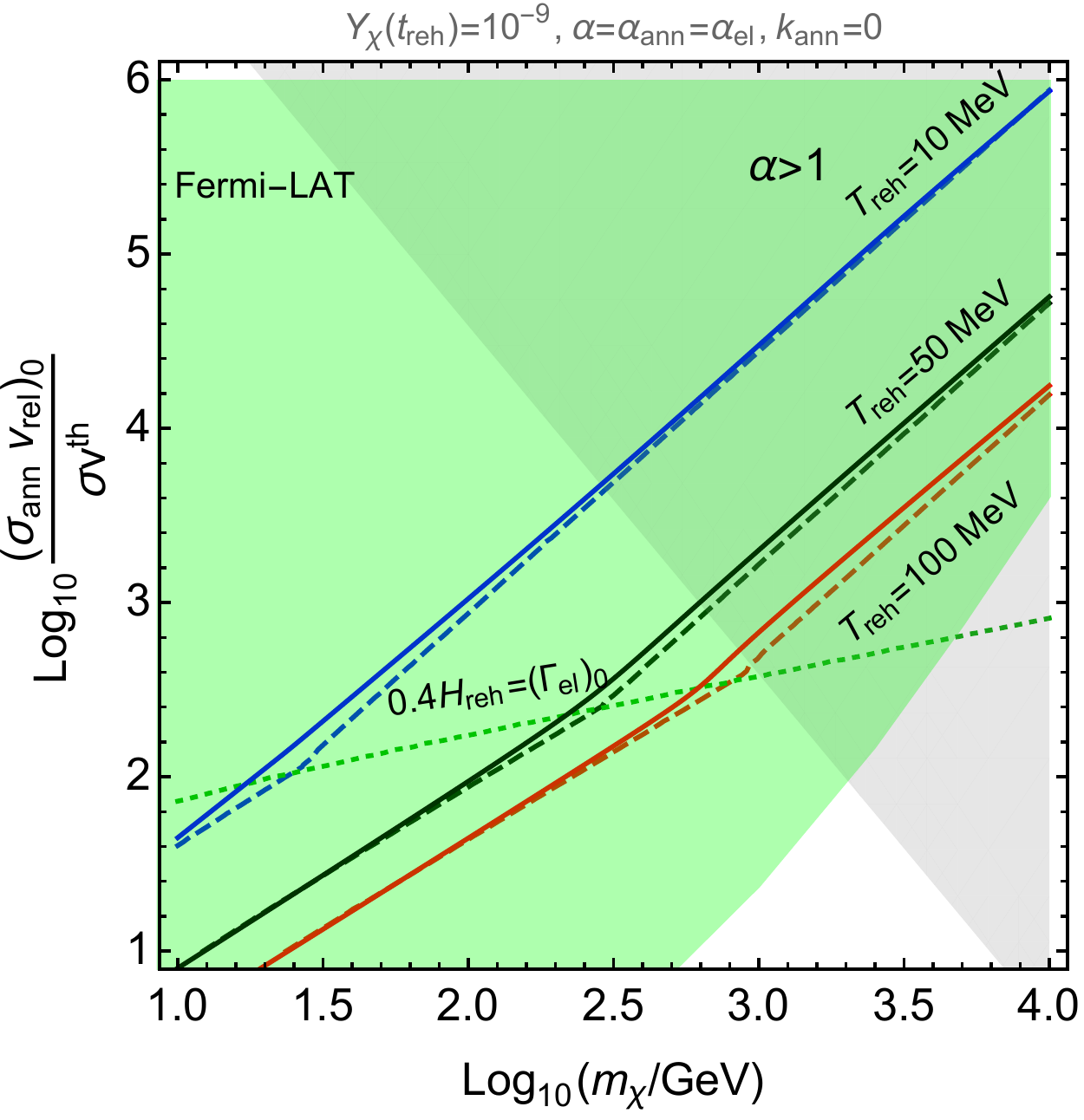} 
    \caption{The contour plot for $k_{\rm ann}=0$, with lines satisfying $\Omega_\chi h^2 = 0.11$ for different $T_{\rm reh}$s;
    $10$\,MeV\,(blue), $50$\,MeV\,(black), $100$\,MeV\,(red) in the plane of dark matter mass and $\langle\sigma_{\rm ann}v_{\rm rel}\rangle_0$.
    $\sigma$v$^{\rm th}\equiv 3\times 10^{-26}{\rm cm}^3/\sec$.
The gray colored region is excluded by conservative perturbative and unitary criterion ($\alpha > 1$). 
Most of regions are already excluded by indirect detection constraints from the Fermi-LAT \cite{Ackermann:2015zua} (green colored region). 
 The dashed lines are for the analytic approximation in Eqs.~(\ref{eq:yield_CA},\ref{eq:yield_IA}).    }     \label{fig:DM_density_s} \end{center}
\end{figure}

\begin{figure}[t]
  \begin{center}
    \includegraphics[width=8.2cm]{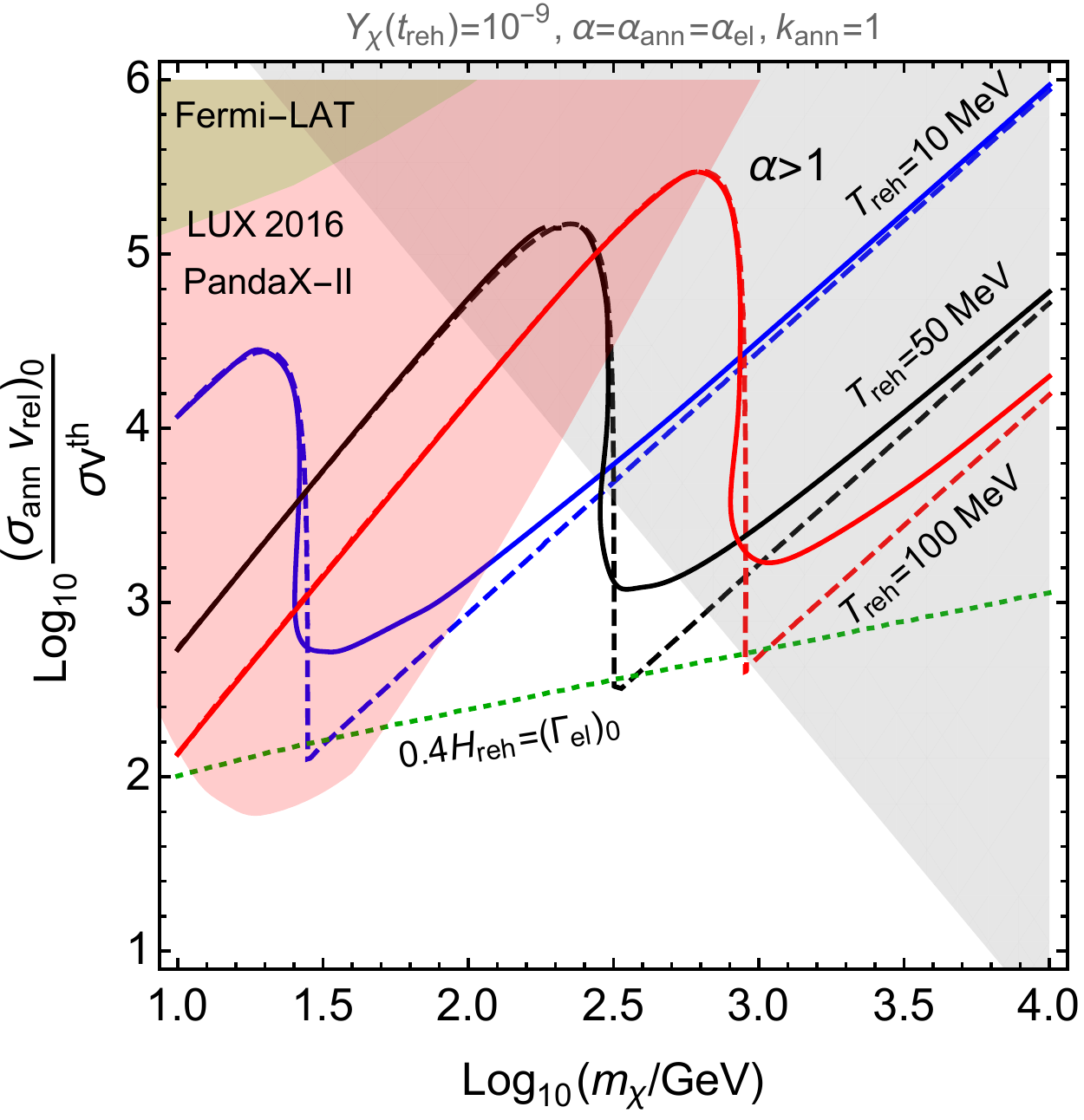}  \\
    \caption{The contour plot for $k_{\rm ann}=1$. 
 The red colored region is excluded by  dark matter direct detection experiment, PandaX-II \cite{Tan:2016zwf} and LUX 2016 \cite{Akerib:2016vxi}.
     }     \label{fig:DM_density_p} \end{center}
\end{figure}

In Fig.~\ref{fig:DM_density_s} and \ref{fig:DM_density_p}, we study the allowed parameter space in the plane of 
$m_\chi-\langle\sigma_{\rm ann} v_{\rm rel}\rangle_0$ for different choices of reheating temperature.
For each figures, the green dotted lines stand for the contour to satisfy the present relic density 
with the condition, $0.4H_{\rm reh} = \langle \Gamma_{\rm el}\rangle_0$. 
For $k_{\rm ann}=0$, 
the present dark matter abundance is proportional to $\langle\sigma_{\rm ann} v_{\rm rel}\rangle_0^{-4/3} T_{\rm reh}^{-7/3}m_\chi^2$ 
in the (C.A.) region that corresponds to the diagonal line above the boundary ($0.4 H_{\rm reh} = \Gamma_{\rm el}$). Below the boundary line, 
the production mechanism is in the (I.A.) region, and the corresponding $\Omega_\chi h^2$ is proportional to $\langle \sigma_{\rm ann} v_{\rm rel}\rangle_0^{-1}T_{\rm reh}^{-1} m_\chi$. Therefore the slope is slightly changed. 

For $k_{\rm ann}=1$, the diagonal line on the right hand side is the same as that of the region (C.A.) with $k_{\rm ann}=0$. However 
there is a drastic change around the boundary. 
The vertical line corresponds to the (I.A.) region where the contribution is dominated by the term $c_0H_{\rm rel}/\langle\Gamma_{\rm el}\rangle_0$ in Eq.~(\ref{eq:yield_IA}). 
Consequently, $\Omega_\chi h^2 \propto  T_{\rm reh}^3/m_\chi^2$, and does not explicit depend on the cross-section. 
For a quite small $H_{\rm rel}/\langle\Gamma_{\rm el}\rangle_0$,  $\Omega_{\rm\chi} h^2$ is proportional to 
$\langle\sigma_{\rm ann} v_{\rm rel}\rangle_0^{-1} T_{\rm reh}^{-2} m_\chi^2$, 
so that the  slope is changed again.
The numerical calculation smooth the analytic lines around the boundary between (C.A.) and (I.A.). 
The more correct boundary line is given as $H_{\rm reh} \simeq \langle\Gamma_{\rm el}\rangle_0$. 


\subsection{Direct/Indirect Detection}
At a low temperature below GeV, the quarks are no longer light degrees of freedom, and the interactions between the dark matter 
and leptons are more crucial to determine the dark matter density. On the other hand, the interaction between the dark matter and quarks are more important for the direct/indirect detection. 
In order to give a rather strong correlation, here we take leptophilic dark matter. 

For $k_{\rm ann}=0$, the present dark matter annihilation is dominated by the $s$-wave contribution, 
which is strongly constrainted by the Fermi-LAT data \cite{Ackermann:2015zua} as in Fig.~\ref{fig:DM_density_s}.
Therefore $k_{\rm ann}=1$ is more viable because the annihilation rate at the present Universe is quite suppressed by the
square of the present dark matter velocity ($v_{\chi0}^2\sim 10^{-6}$) compared to $\langle\sigma_{\rm ann} v_{\rm rel}\rangle_0$.

For the direct detection of the dark matter, the 1-loop or 2-loop induced
interaction between the dark matter and nucleus can generate the sizable signal. 
 One can think various effective operators for $\chi\chi l l$ 
  with complex couplings and various spin structure.  
As a benchmark example,  we assume that the dark matter is a Majorana fermion, 
and that the interaction is mediated by a real scalar. 
Using four component spinor notation,  the relevant effective operator is given as
\beq \label{eq:eff_op}
{\cal L}_{\rm eff}= \frac{(\bar\chi\chi)(\bar l \emph{l})}{\Lambda^2}.
\eeq 
We consider lepton flavor universal couplings in order not to generate any flavor problem. 
After matching the effective operator of Eq.~(\ref{eq:eff_op}) to that 
for the scattering cross-section of Eq.~(\ref{eq:scat_op}), 
we can apply the constraints from direct detection experiments \cite{Tan:2016zwf,Akerib:2016vxi}.

For Eq.~(\ref{eq:eff_op}), the first non-vanishing dark matter nucleus ($N$) elastic scattering cross-section is generated at the two-loop level; \cite{Kopp:2009et}
\beq
\sigma_{\chi N} =& {\cal O}(1) 
 \left(\frac{\mu_N^2 Z^2}{\pi} \right)\left(\frac{e^4 Z}{192\pi^2 \Lambda^2}\right)^2 
 \left(\frac{\mu_N v_{\chi 0}}{m_l}\right)^2\nonumber\\
 \equiv& \frac{ \mu_N^2}{\mu_n^2} A^2 \sigma_{\chi n},
\eeq  where $n$ is the nucleon. 
The ${\cal O}(1)$ uncertainties are coming from the two-loop induced 
nucleus form factor, whose evaluation is beyond the scope of this paper. 
$Z$ is the atomic number, $A$ is the mass number of the target nucleus.
$Z=54,\, A=131$ for $^{131}$Xe. 
$\mu_N = m_\chi m_N/(m_\chi+m_N)$ 
is the reduced mass for the dark matter and the nucleus, and  
$\mu_n$ is the reduced mass for the the dark matter and a nucleon. 
$\mu_N v_{\chi_0}$ is the typical recoil momentum of the nucleus, and 
the formula is valid for $\mu_N v_{\chi_0} ={\cal O}({\rm MeV}) \lesssim m_l$. 
Taking all those uncertainties as a factor ${\cal O}(1)$,  in Fig.~\ref{fig:DM_density_p}
we get the excluded region of the cross-section (red color) for a given dark matter mass.
Even though it is generated at a two-loop level,  the strong constraint exists for the range that satisfies the dark matter density. More accurate constraints considering all the ${\cal O}(1)$ coefficients correctly will be discussed in future work.

\section{Conclusions}\label{sec:Conc}

Non-thermal history of the early Universe can be naturally obtained in new physics beyond the Standard Model, and it also provides various interesting effects which cannot be simply captured by the standard thermal history of the Universe with a high reheating temperature, including the works \cite{Arcadi:2011ev,Hooper:2011aj,Dev:2013yza,Kamada:2014ada}. 

In this work, we have studied the effect of the elastic scattering 
between the WIMP dark matter and background 
radiations when the Universe is reheated at a low temperature. 
This effect is crucial if the amount of non-thermally produced dark matter 
is sizable, and the reheating temperature is well below the freeze-out temperature. 
We specified the three conceptual domains for the determination of 
the dark matter abundance, and 
presented the analytic and numerical solutions to the Boltzmann equation. 

When the reheating temperature is low enough, the elastic scattering rate 
is not effective to completely thermalize the dark matter.
The dark matter particles decouple from thermal plasma when they are still relativistic, and the annihilation could persist until they become non-relativistic. In this case, we show that the final abundance of the dark matter could depend on the elastic scattering rate. Even in the case of instantaneous thermalization,
the relative size between the elastic and annihilation rates can change
the final abundance for the $p$-wave annihilating dark matter. 
On the other hand, the non-thermal WIMP mechanism requires large annihilation cross-section to explain the present dark matter relic density. We studied the constraints from direct/indirect detection experiments by considering the leptophilic dark matter model as a specific example, and showed that wide range of parameter space is severely constrained.

Those strong constraints can be avoided if the dark matter is ``Dark WIMP"
in which the dark matter is thermalized by and annihilates to dark radiations. 
 The mechanisms that we have discussed can also be generalized to the dark WIMP scenario.
In such a case, there could be more interesting connection between 
the history of the early Universe and the signatures imprinted on the cosmic microwave background and large scale structure. \\

\section*{Acknowledgements}
We would like to thank Jinn-Ouk Gong, Kyu Jung Bae, Jong-Chul Park, and Seodong Shin for useful discussions. HK is supported by IBS under the project code, IBS-R018-D1. JH is supported by World Premier International Research Center Initiative (WPI Initiative), MEXT, Japan. CSS is supported in part by the Ministry of Science, ICT\&Future Planning and by the Max Planck Society, Gyeongsangbuk-Do and Pohang City.

\begin{appendix}
\section{The Boltzmann equation for the momentum expectation value of dark matter}\label{apdix}
In this appendix, we derive the full Boltzmann equation for the momentum expectation value 
of dark matter including the annihilation effect. This effect could be important because  the annihilation cross-section depends on the momentum, so the dark matter with different momentum will annihilate with a different rate. 

The Boltzmann equation for the distribution is 
\beq\label{eq:Botlz}
\frac{ d f_\chi (\vec p)}{d t}  =  C_{\rm el} [f_\chi]  + C_{\rm ann}[f_\chi].
\eeq
The collision terms are given as
\beq
C_{\rm el}[f_\chi]
= &\frac{1}{E_p} \int d \Pi_q d \Pi_{p'} d\Pi_{q'} (2\pi)^4
\delta^{(4)}(p_\mu+q_\mu - p_\mu'-q_\mu')\nonumber\\
& \hskip 1cm |{\cal M}_{\rm el}|^2 
[ f_\chi (\vec p') f_\gamma ( \vec q') - f_\chi(\vec p) f_\gamma(\vec q)],\nonumber\\
C_{\rm ann}[f_\chi] 
= & \frac{1}{E_p}\int d\Pi_{p'} d\Pi_{q} d\Pi_{q'}(2\pi)^4 
\delta^{(4)}(p_\mu+p_\mu'-q_\mu-q_\mu')\nonumber\\
&\hskip 1cm |{\cal M}_{\rm ann}|^2 
[ f_\gamma(\vec q)f_\gamma(\vec q') - f_\chi(\vec p) f_\chi (\vec p')],
\eeq
where $d\Pi_{p_i} = (2\pi)^{-3}(2 E_{p_i})^{-1}d^3\vec p_i$. Integrating Eq.~\eqref{eq:Botlz} over the phase space of $\chi$, we could obtain equations for dark matter number density and momentum expectation value. Assuming $n_\chi \gg n_\chi^{\rm eq}$, the equation for number density is 
\beq\label{eq:Boltz_num}
\dot n_\chi + 3 H n_\chi =& - \langle \sigma_{\rm ann} v_{\rm rel}\rangle_{\chi} n_\chi^2 \nonumber\\
\equiv &  -\Gamma_{\rm ann} n_\chi,
\eeq
and the equation for momentum expectation value is
\beq\label{eq:Boltz_full_mom}
\dot {\langle p\rangle}_\chi
+ H \langle p\rangle_\chi 
=&  -\langle \sigma_{\rm el} v_{\rm rel} \Delta p\rangle_{\chi  T} n_\gamma 
+ \langle \sigma_{\rm ann} v_{\rm rel}\rangle_{\chi}\langle p\rangle_\chi n_\chi  
\nonumber\\
&\,  - \langle\sigma_{\rm ann} v_{\rm rel} p\rangle_{\chi} n_\chi  \nonumber\\
\equiv & - (\Gamma_{\rm el}  - \Gamma_{\rm ann} S)\langle p\rangle_\chi .
\eeq
Here $p=|\vec p|$, and $\Delta p$ is the change of dark matter momentum for single elastic scattering. 
$S$ is defined as 
\beq
S =  1 - \frac{\langle \sigma_{\rm ann} v_{\rm rel} p\rangle_{\chi}}{\langle p\rangle_\chi \langle\sigma_{\rm ann} v_{\rm rel}\rangle_{\chi}}.
\eeq
It is obvious that the annihilation affects the momentum evolution, and it relys on $\Gamma_{\rm ann}S$.

It is difficult to obtain an exact analytic form of $S$ for arbitrary distribution functions. But it is possible to approximate $S$ as
\beq
S \simeq \frac{\sigma_{sd,\chi}^2}{\langle p \rangle_\chi^2},
\eeq
when the distribution function have small variance compared to its mean value; $\sigma_{sd, \chi}^2  =  \langle p^2\rangle_\chi - \langle p\rangle_\chi^2 \ll \langle p \rangle_\chi^2 $.


At the end of reheating, dark matter particles are produced from the decay of long-lived heavy particle, $\phi$. If they are produced by two-body decays, the dark matter momentum would be centered at $\langle p \rangle_\chi \simeq m_\phi/2$, and the variance (or width) of distribution would be given as $\sigma_{sd,\chi}^2 \simeq \Gamma_\phi^2$. Thus, we see that $S(t_{\rm reh}) \simeq ( \Gamma_\phi / m_\phi)^2 \ll 1$, and that the annihilation does not change the momentum evolution as long as $S < \Gamma_{\rm el}/\Gamma_{\rm ann}$. 

This argument is only valid at the end of reheating because the elastic scattering spreads the momentum distribution of dark matter. To make sure that we can safely ignore the annihilation of dark matter for its momentum evolution, it is necessary to investigate the time evolution of $S$. 


Let us consider a case where $ H < \Gamma_{\rm el} <\Gamma_{\rm ann} $ at $t= t_{\rm reh}$. This is the case where the annihilation may play a role in determining momentum evolution. In other cases, the effect of $\Gamma_{\rm ann}$  on momentum evolution is safely ignored. Just after dark matters are produced, the width will evolve dominantly by elastic scatterings. When $\sigma_\chi$ is still smaller than $\langle p\rangle_\chi$, it is straightforward to derive the Boltzmann equation for $S$. It is
\beq
\frac{d S}{dt}  = & 2 \Gamma_{\rm el} S  + \Gamma_{\rm el} 
\frac{\langle\Delta p\rangle_\chi}{\langle p\rangle_\chi}  
+ {\cal O}(S^2).
\eeq
For a small initial value of $S(t_{\rm reh}) = (\Gamma_\phi/m_\phi)^2$, the second term of RHS becomes the source of $S$. Then solution becomes
\beq
S\simeq  \Delta t\, \Gamma_{\rm el} \frac{\langle\Delta p\rangle_\chi}{\langle p\rangle_\chi},
\eeq
where $ \Delta t \equiv t-t_{\rm reh} < 1/\Gamma_{\rm el} $. On the other one hand, 
for $1/\Gamma_{\rm ann}|_{T_{\rm reh}}< \Delta t < H_{\rm reh}^{-1}$, 
the solution to Eq.~\eqref{eq:Boltz_num} is
 $n_\chi \simeq (\langle\sigma_{\rm ann} v_{\rm rel} \rangle_{\chi\chi}\Delta t)^{-1}$ 
 so that $\Gamma_{\rm ann} \simeq 1/\Delta t$.
Inserting these solutions to Eq.~\eqref{eq:Boltz_full_mom} gives 
\beq
\dot{\langle p\rangle}_\chi + H\langle p\rangle_\chi \simeq 
- \left(\Gamma_{\rm el}  - \Gamma_{\rm el}\frac{\langle \Delta p\rangle_\chi}{\langle p\rangle_\chi}\right)\langle p\rangle_\chi.\eeq
When the dark matter is non-relativistic at the center of mass frame for $\chi-\gamma$ collision system, $\Delta p/p \ll 1$. This means that the momentum evolves dominantly by elastic scattering until the relaxation time, $\Delta t\sim 1/\Gamma_{\rm el}$.  


\end{appendix}

\bibliography{references}

\end{document}